\documentclass[twocolumn,prd,showpacs]
              {revtex4}

\usepackage{graphicx}

\begin{document}

\title{Dynamics of massive matter disappearance on the brane}

\author{Michael~Maziashvili}

\email{maziashvili@ictsu.tsu.edu.ge} \affiliation{Department of
Theoretical Physics, Tbilisi State University,\\ 3 Chavchavadze
Ave., Tbilisi 0128, Georgia \\Institute of High Energy Physics and
Informatization,\\ 9 University Str., Tbilisi 0186, Georgia }

\begin{abstract}
Time evolution of the probability of quasilocalized massive modes
decay trough the tunneling into an extra dimension is analyzed in
the Randall-Sundrum braneworld model.
\end{abstract}

\pacs{11.10.Kk}


\maketitle
\section{Introduction}
The idea that our world is a three brane embedded in a higher
dimensional space-time with non-factorizable warped geometry has
been much investigated since the appearance of papers \cite{Go,
Ra}. The first solution for a warped metric was provided in
\cite{Go}. In this paper it was suggested that gravitational
interactions between particles on a brane in uncompactified
five-dimensional space could have the correct four-dimensional
Newtonian behavior, provided that the bulk cosmological constant
is fine-tuned to the brane tension. As it was shown in \cite{Ra}
the warped metrics could provide an alternative to
compactification of extra dimension used for a possible
explanation of the hierarchy problem \cite{AADD}. Brane solution
described in \cite{Go, Ra} has the form
\begin{equation}\label{GRS}ds^2=e^{-2\kappa|z|}\eta_{\mu\nu}dx^{\mu}dx^{\nu}-dz^2~,\end{equation}
where the parameter $k$ is determined by the bulk cosmological and
five-dimensional gravitational constants respectively. An
important issue in the study of braneworld models is the
localization of standard model fields on the brane. This problem
for the brane (\ref{GRS}) was addressed in \cite{BG}. It was shown
that the massless scalar field has a zero mode normalizable with
the weight $\exp(-2k|z|)$. When this field is given mass the bound
state becomes metastable against tunneling into an extra dimension
\cite{DRT}. Fermion fields are not localized on the positive
tension brane by gravitational interactions \cite{BG}. In order to
localize fermions one can use the method of localization described
in paper \cite{JR}. But the adding of bulk mass term makes the
fermion bound state unstable as well trough the barrier
penetration into an extra dimension \cite{DRT}. Based on this
phenomenon the possibility of decaying cold dark matter into an
extra space was considered in the framework of braneworld
cosmology, see \cite{IGKMY}. The comoving density of cold dark
matter was taken to decay over time with a rate $\rho
a^3\exp(-\Gamma t)$ corresponding to the exponential decay of a
metastable state. But in Quantum Mechanics it is well established
that an exponential decay cannot last forever if the Hamiltonian
is bounded below and cannot occur for small times if, besides
that, the energy expectation value of the initial state is finite
\cite{CSMFGR}. From this point of view we consider the dynamics of
massive matter disappearance on the brane.

\section{Scalar field} Due to four-dimensional Poincare invariance
of the metric (\ref{GRS}) every fields in this background can be
decomposed into four-dimensional plane waves

\begin{equation}\label{plwde}\phi\propto \exp(ip_{\mu}x^{\mu})\varphi(z)~,\end{equation} where
coordinate four momentum $p_{\mu}$ coincides with the physical
momentum on the brane. It is natural to assume that the three
momentum $\vec{p}$ does not change the picture of tunneling into
an extra dimension. For the sake of simplicity we put $\vec{p}=\bf
0$ in what follows. Under this assumption the equation

\begin{equation}\label{timequat}\partial^2_t\phi-e^{2k|z|}\partial_z(e^{-4k|z|}\partial_z\phi)+e^{-2k|z|}\mu^2\phi=0~,\end{equation}
gives

\begin{equation}\label{eigeq}-\partial^2_z\varphi+4k\,\mbox{sgn}(z)\partial_z\varphi+\mu^2\varphi=
e^{2k|z|}E^2\varphi~,\end{equation} where $E\equiv p_{0}$. The
continuum spectrum of eq.(\ref{eigeq}) starts from $E^2=0$

\[\varphi=\sqrt{{E\over 2k}}e^{2k|z|}\left[a(E)J_{\nu}\left({E\over k}e^{k|z|}\right)+b(E)Y_{\nu}\left({E\over k}e^{k|z|}\right)\right]~,\]
where $\nu=\sqrt{4+\mu^2/k^2}$ \cite{DRT, Ru}. The normalization
condition
\[\int\limits_{-\infty}\limits^{\infty}dze^{-2k|z|}\varphi(E,~z)\varphi(E',~z)=\delta(E-E')~,\]
as well as the boundary condition $\partial_z\varphi|_{z=0}=0$ can
be satisfied by taking \cite{DRT}

\[a(E)=-{A(E)\over \sqrt{1+A^2(E)}}~,~~b(E)={1\over \sqrt{1+A^2(E)}}~,\]

\[A(E)={Y_{\nu-1}(E/k)-(\nu-2)(k/E)Y_{\nu}(E/k)\over J_{\nu-1}(E/k)-(\nu-2)(k/E)J_{\nu}(E/k)}~.\]

The general solution to Eq.(\ref{timequat}) can be written as
\cite{Lee}

\begin{eqnarray}\label{gensol}\phi(t,~z)&=&\int\limits_{-\infty}\limits^{\infty}G_1(t,~z,~z')\phi(0,~z')dz'\nonumber\\
&+&\int\limits_{-\infty}\limits^{\infty}G_2(t,~z,~z')\dot{\phi}(0,~z')dz'~,\end{eqnarray}
where
\begin{eqnarray}G_1(t,~z,~z')=e^{-2k|z'|}\int\limits_0\limits^{\infty}dE\cos(Et)\varphi(E,~z)\varphi(E,~z')~,\nonumber\\
G_2(t,~z,~z')=e^{-2k|z'|}\int\limits_0\limits^{\infty}dE{\sin(Et)\over
E}\varphi(E,~z)\varphi(E,~z')~.\nonumber\end{eqnarray} To compute
$\phi(t,~z)$ one needs to know the initial data
$\phi(0,~z),~\dot{\phi}(0,~z)$. In what follows we take the
following condition $\dot{\phi}(0,~z)=iE_0\phi(0,~z)$
corresponding to the free particle on the brane (\ref{plwde}) with
energy $E_0$. Correspondingly the Eq.(\ref{gensol}) takes the form
\begin{equation}\label{eveq}\phi(t,~z)=\int\limits_{-\infty}\limits^{\infty}\left[G_1(t,~z,~z')+iE_0G_2(t,~z,~z')\right]
\phi(0,~z')dz'~.\end{equation} At $t=0$ the particle is confined
on the brane. The probability of the particle to remain on the
brain at instant $t$ is given by
\[\left|\left<\phi(0)|\phi(t)\right>\right|^2~,\] where $\phi(0)$ denotes the brane localized initial state and the scalar product
is understood with the measure $\exp(-2k|z|)$
\[\left<\phi(0)|\phi(t)\right>=\int\limits_{-\infty}^{\infty}dze^{-2k|z|}\phi^*(0,~z)\phi(t,~z)~.\] From Eq.(\ref{eveq}) one finds
\begin{equation}\label{realpart}Re\left<\phi(0,~z)|\phi(t,~z)\right>=Re\left[\int\limits_0\limits^{\infty}dEe^{-iEt}\left|C(E)\right|^2\right]~,\end{equation}
\begin{equation}\label{imaginpart}Im\left<\phi(0,~z)|\phi(t,~z)\right>=-E_0Im\left[\int\limits_0\limits^{\infty}dE{e^{-iEt}\over E}\left|C(E)\right|^2\right]~,\end{equation}
where
\[C(E)=\left<\phi(0)|\varphi(E)\right>~.\]

\begin{figure}
  \begin{center}

\includegraphics{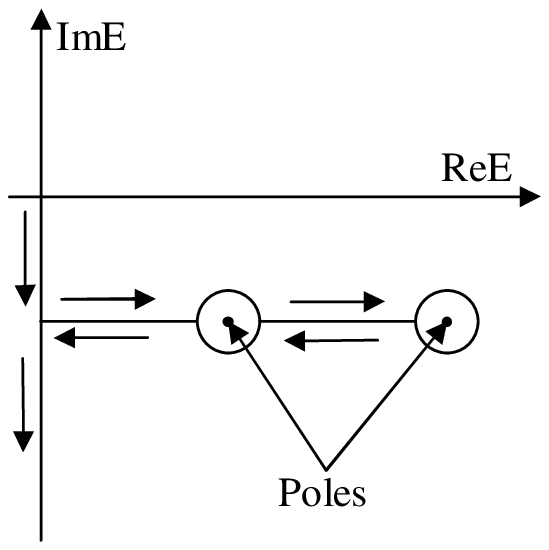}

 \end{center}

  \caption{The integration contour.}

\end{figure}

In the above $E$ is real variable, but in the following it is
understood that $E$ can be complex. The function
$\left|C(E)\right|^2$ has the simple poles determined by the
equation \[i\pm A(E)=0~.\] The equation
\[A(E)=i~\Rightarrow~{E\over k}H^{(1)}_{\nu-1}\left({E\over k}\right)+(2-\nu)H^{(1)}_{\nu}\left({E\over k}\right)=0~,\]
under assumption $\mu/k, E/k\ll 1$ gives the pole in the fourth
quadrant of a complex $E$ plane, $E=E_0-i\Gamma$, where
\[E_0={\mu\over \sqrt{2}}~,~~~{\Gamma\over E_0}={\pi\over
8}\left({E_0\over k}\right)^2~.\] Since, when $t > 0$ and
$|E|\rightarrow\infty$, $e^{-iEt}\rightarrow 0$ in the fourth
quadrant, for evaluating of integrals entering
Eqs.(\ref{realpart},~\ref{imaginpart}) one can deform the
integration counter as it is shown in Fig.1. In this way one finds
\begin{eqnarray}\int\limits_0\limits^{\infty}dEe^{-iEt}\left|C(E)\right|^2 &=&\nonumber\\
 \mbox{residue term}&-&i\int\limits_0\limits^{\infty}
dEe^{-Et}\left|C(-iE)\right|^2.\end{eqnarray} From this equation
one sees that the second term gives an imaginary contribution and
correspondingly $Re\left<\phi(0,~z)|\phi(t,~z)\right>$ is
determined by the residue term only. In the same way one finds
that $Im\left<\phi(0,~z)|\phi(t,~z)\right>$ is also determined
only by the residue term. Following to the quantum mechanical
formalism, and by taking into account that $\mu/k,~E_0/k\ll 1$ and
the localization width of the scalar is $\sim k^{-1}$, one can
simply take
\begin{equation}\label{initstate}\phi(0,~z)=\left\{\begin{array}{ll}\sqrt{k}/\sqrt{1-e^{-1}}~, & \mbox{for}~~ |z|\leq (2k)^{-1}~,\\
0~, & \mbox{for}~~ |z|>(2k)^{-1}~.
\end{array}\right.\end{equation} Thus, the probability of finding the particle on the brane equals one when $t=0$. By
evaluating the residue term and using Eq.(\ref{initstate}) one
finds

\[\left|\left<\phi(0)|\phi(t)\right>\right|^2=0.04\times e^{-2\Gamma t}~.\]

\section{Spin 1/2 field} The minimal representation of spinors in
five dimensions can be chosen to be four dimensional. The
five-dimensional Minkowskian gamma matrices can be chosen as
follows

\[\begin{array}{ll}\Gamma^{\mu}=\gamma^{\mu}~,\\\Gamma^z=-i\gamma^5~,\end{array}~\gamma^{\mu}=\left(\begin{array}{cc}
0 & \sigma^{\mu}\\ \bar{\sigma}^{\mu} & 0
\end{array}\right)~,~\gamma^5=\left(\begin{array}{cc}
1 & 0 \\ 0 & -1
\end{array}\right)~,\] where $\sigma^{\mu}=({\bf 1}~,~\vec{\sigma})~,~\bar{\sigma}^{\mu}=({\bf
1}~,~-\vec{\sigma})$ with $\sigma_i$ the three Pauli matrices. The
Dirac equation in the background
\[ds^2=e^{-2\sigma(z)}\eta_{\mu\nu}dx^{\mu}dx^{\nu}-dz^2~,\]
for the fermion coupled with the domain wall $\Phi$ reads

\begin{equation}\label{fermeq}\left[i\Gamma^z(\partial_z-2\sigma'(z))+ie^{\sigma(z)}\Gamma^{\mu}\partial_{\mu}-g\Phi\right]\psi=0~.\end{equation}
Solutions to this equation are linear combination of wave
functions of the form \[\psi=\exp(-ip_{\mu}x^{\mu})\chi~.\] We do
not expect the three momentum $\vec{p}$ to affect the overall
picture of particle tunneling into an extra space. Therefore, as
in the case of scalar field, we assume $\vec{p}=\bf 0.$ Under this
assumption the Eq.(\ref{fermeq}) takes the form
\begin{equation}\label{fereqimzer}\left(i\partial_0-H\right)\psi=0~,\end{equation}
where
\begin{equation}\label{ferhamimzer}H=e^{-\sigma}\left[\gamma^0g\Phi-\gamma^0\gamma^5(\partial_z-2\sigma')\right]~.\end{equation}
So that the time evolution of initially brane localized fermion
$\psi(0,~z)$ is given by
\begin{equation}\psi(t,~z)=\int\limits_{0}\limits^{\infty}dEe^{-iEt}\chi(E,~z)\left<\chi(E)|\psi(0)\right>~,\end{equation}
where the scalar product is understood with the measure
$e^{-3k|z|}$ and
\begin{equation}\label{fermeigeneq}H\chi=E\chi~.\end{equation}
In terms of the right- and left-handed components
\[\chi_R\equiv {1+\gamma^5\over 2}\chi~,~~~~\chi_L\equiv
{1-\gamma^5\over 2}\chi~,\] the Eq.(\ref{fermeigeneq}) takes the
form

\begin{equation}\label{chiralll}\left(\partial_z-2\sigma'-g\Phi\right)\chi_R=-Ee^{\sigma}\chi_L~,\end{equation}

\begin{equation}\label{chirallr}\left(\partial_z-2\sigma'+g\Phi\right)\chi_L=Ee^{\sigma}\chi_R~,\end{equation}
where $E\equiv p_0$. After eliminating $\chi_R$ from
Eqs.(\ref{chiralll},~\ref{chirallr}) one obtains a second order
equation for $\chi_L$
\begin{eqnarray}\left[\partial_z^2-5\sigma'\partial_z-2\sigma''+6(\sigma')^2-g\sigma'\Phi\right.\nonumber\\
\left.+g\Phi'-(g\Phi)^2+E^2e^{2\sigma}\right]\chi_L=0~.\end{eqnarray}
For the domain wall profile in the thin wall limit one can take
\[\Phi=v\,\mbox{sgn}(z)~.\]
To the right ($z>0$) and left ($z<0$) of the brane  one gets
\begin{eqnarray}\label{leftrightz}\left[\partial_z^2-5k\partial_z-gkv+6k^2-g^2v^2+E^2e^{2kz}\right]\chi_L=0~,\nonumber\\
\left[\partial_z^2+5k\partial_z-gkv+6k^2-g^2v^2+E^2e^{-2kz}\right]\chi_L=0~.
\end{eqnarray} The solution of Eq.(\ref{leftrightz}) reads
\begin{equation}\chi_L=\sqrt{{Ee^{5k|z|}\over 2k}}\left[a(E)J_{\nu}\left({E\over
k}e^{k|z|}\right)+b(E)Y_{\nu}\left({E\over
k}e^{k|z|}\right)\right]~,\end{equation} where $\nu=(2gv+k)/2k$.
From Eqs.(\ref{chiralll},\ref{chirallr}) one finds the following
boundary condition
\[\chi_L'(0)+(gv-2k)\chi_L(0)=0~,\] where the prime stands for the derivative with respect to $|z|$.
To satisfy this boundary condition as well as the normalization
condition
\[\int\limits_{-\infty}\limits^{\infty}dze^{-3k|z|}\chi_L(E,~z)\chi_L(E',~z)=\delta\left(E-E'\right)~,\]
the coefficients $a(E),~b(E)$ should have the form \cite{Tit}
\[a(E)={Y_{\nu-1}(E/k)\over \sqrt{Y^2_{\nu-1}(E/k)+J^2_{\nu-1}(E/k)}}~,\]\[b(E)=-\sqrt{1-a^2(E)}~.\]
The solution to Eq.(\ref{leftrightz}) for $E=0$
\begin{equation}\label{zeroenfersol}\psi_L\,\propto\,\,\, e^{\left(2k-gv\right)|z|}~,\end{equation}
is localized on the brane as long as $k<2gv$ with the localization
width $2/(2gv-k)$ \cite{BG}. Under assumption that the initial
state is given by the left-handed particle localized on the brane
one finds
\begin{equation}\label{leamlitu}\left<\psi_L(0)|\psi_L(t)\right>=\int\limits_0\limits^{\infty}dEe^{-iEt}\left|\left<\chi_L(E)|\psi_L(0)\right>\right|^2~.\end{equation}
The function $\left|\left<\chi_L(E)|\psi_L(0)\right>\right|^2$ may
have the poles $E_n$ in the fourth quadrant of complex $E$ plane
determined by the zeros of
$H^{(1)}_{\nu-1}(E/k)H^{(2)}_{\nu-1}(E/k)$ indicating the presence
of resonances in the spectrum of KK modes. For evaluating the
integral (\ref{leamlitu}) one can deform the integration contour
as it is shown in Fig.1. In this way one gets
\begin{eqnarray}\label{fermintdefor}&\left<\psi_L(0)|\psi_L(t)\right>=\nonumber\\
&\sum\limits_nc_ne^{-iE_nt}-i\int\limits_0\limits^{\infty}dEe^{-Et}\left|\left<\chi_L(-iE)|\psi_L(0)\right>\right|^2~,\end{eqnarray}
where \[c_n=-2\pi i\lim\limits_{E\rightarrow
E_n}(E-E_n)\left|\left<\chi_L(E)|\psi_L(0)\right>\right|^2~.\] So
that the probability to find the particle on the brane at time $t$
is given by
\[\left|\left<\psi_L(0)|\psi_L(t)\right>\right|^2=\sum\limits_n|c_n|^2e^{2Im(E_n)t}+\mbox{interference terms}~.\]
One can easily check that
\[\left.{d\left<\psi_L(0)|\psi_L(t)\right>\over
dt}\right|_{t=0}=0~.\] Namely, by using
Eqs.(\ref{fereqimzer},\ref{ferhamimzer}) and taking into account
that $\phi(0,~z)$ is an even function of $z$ one sees that in this
expression the odd function is integrated over a symmetric region.
So that the evolution of the decay process requires some time to
reach the regime of exponential decay. For large values of time
$t\gg k^{-1}$ the second term in Eq.(\ref{fermintdefor}) becomes
dominant for which in this limit only modes with $E\ll k$ are
relevant. Therefore, the non-escape probability behaves
asymptotically as
\[\left|\left<\psi_L(0)|\psi_L(t)\right>\right|^2\sim (kt)^{-4\nu+4}~.\]
Thus, after a long time the decay proceeds with respect to the
power law. As it is shown in \cite{CS} one can choose the index
$\nu$ in such a way ($2< \nu < 5$) the Hankel function
$H^{(1)}_{\nu-1}(\omega)$ to have exactly one zero in the fourth
quadrant of complex $\omega$ plane. This is the only zero of the
product $H^{(1)}_{\nu-1}(\omega)H^{(2)}_{\nu-1}(\omega)$ located
in the fourth quadrant since for real values of $\nu$,
$H^{(1)}_{\nu}(\omega)^*=H^{(2)}_{\nu}(\omega^*)$. From the
trajectories of zeros presented in \cite{CS} one sees that if
$\nu$ is close to $2.5$ ($\nu > 2.5$) the function
$H^{(1)}_{\nu-1}(E/k)$ has the zero $E=E_0-i\Gamma$ in the fourth
quadrant such that $E_0/k\ll 1$ and $\Gamma\sim k$. Using Eq.(
\ref{zeroenfersol}) for the initial state of the low lying KK
resonances one can simply take
\begin{equation}\label{ferinitstate}\psi_L(0)=\left\{\begin{array}{ll}{\sqrt{2gv-k
\over 2(1-e^{-1})}}e^{(2k-gv)|z|}~, & ~~ |z|\leq (2gv-k)^{-1}~,\\
0~, & ~~ |z|>(2gv-k)^{-1}~.
\end{array}\right.\end{equation} So that at $t=0$ the particle is
known to be on the brane, $|z|\leq (2gv-k)^{-1}$, with probability
one. Using Eq.(\ref{ferinitstate}) and the previous consideration
one can perform the concrete calculations for different low lying
KK resonances.

\section{Discussion and Conclusion} From the present consideration
one sees that if the transverse equation contains the second order
time derivative the corresponding decay law has the exponential
form as long as $\Gamma\ll E_0$. So that under this assumption the
decay of massive initially brane localized metastable modes for
scalar, vector and gravitational fields follows to the exponential
law. It should be stressed that the choice
$\dot{\phi}(0,~z)=iE_0\phi(0,~z)$ which is essential to this
result is not unique. However we find this condition to be natural
as it corresponds to the free particle on the brane (\ref{plwde})
with energy $E_0$ corresponding to the metastable state. In
contrast, the decay of massive quasilocalized fermion modes does
not follow completely to the exponential law. As in the standard
quantum mechanical case mentioned in the introduction, initially
the decay is slower than exponential, then comes the exponential
region and after a long time it obeys a power law. The present
approach allows one to evaluate the preexponential term as well as
the whole dynamics of the decay and can be important to evaluate
dynamics for any process based on this phenomenon, for instance to
examine more precisely the cosmological constraints on the
paradigm of decaying dark matter in the braneworld model.

\subsection*{Acknowledgements} The author is greatly indebted
to Z.~Berezhiani, M.~Gogberashvili and I.~Gogoladze for useful
conversations. The work was supported by the grant FEL. REG.
$980767$.


\end{document}